      [1999/12/01 v1.4c Il Nuovo Cimento]
\documentclass{cimento}

\def\aa{{\em A\&A}\ }

\def\aj{{\em AJ}\ }

\def\apj{{\em ApJ}\ }
\def\apjs{{\em ApJS}\ }

\def\nat{{\em Nature}\ }

\def\lsim{\mathrel{\rlap{\lower 4pt \hbox{\hskip 1pt $\sim$}}\raise 1pt
\hbox {$<$}}} 
\def\gsim{\mathrel{\rlap{\lower 4pt \hbox{\hskip 1pt $\sim$}}\raise 1pt
\hbox {$>$}}}

\newcommand{\eg}{e.g., }
\newcommand{\ie}{i.e., }
\newcommand{\Msun}{M_{\odot}}

\newcommand{\kms}{km~s$^{-1}$}

\newcommand{\OI}{O~{\sc i}}
\newcommand{\SiII}{Si~{\sc ii}}
\newcommand{\CaII}{Ca~{\sc ii}}
\newcommand{\FeII}{Fe~{\sc ii}}
\newcommand{\FeIII}{Fe~{\sc iii}}
\newcommand{\TiII}{Ti~{\sc ii}}

\newcommand{\Nifs}{$^{56}$Ni}
\newcommand{\Mej}{M_{\rm ej}}

\newcommand{\Mni}{M{\rm (^{56}Ni)}}
\newcommand{\Ed}{\dot{E}_{\rm dep}}

\newcommand{\Edep}{\dot{E}_{\rm dep,51}}
\usepackage{graphicx}  

\title{Diversity of the Supernova - Gamma-Ray Burst Connection}
\author{
K.~Nomoto\from{ins:ut}\ETC,
N.~Tominaga\from{ins:ut},
M.~Tanaka\from{ins:ut},
K.~Maeda\from{ins:ko},
T.~Suzuki\from{ins:ut},
J.S.~Deng\from{ins:ch}\from{ins:ut},
\atque
P.A.~Mazzali\from{ins:mpa}\from{ins:tri}\from{ins:ut}
}

\instlist{
\inst{ins:ut} Department of Astronomy, University of Tokyo, Bunkyo-ku, Tokyo
113-0033, Japan
\inst{ins:ko} Department of Earth Science and Astronomy, College of Arts and
Science, University of Tokyo, Meguro-ku, Tokyo 153-8902, Japan
\inst{ins:ch} National Astronomical Observatories, CAS, Beijing 100012, China
\inst{ins:mpa} Max-Planck-Institut f\"ur Astrophysik, 
85741 Garching, Germany
\inst{ins:tri} Istituto Nazionale di Astrofisica-OATs, Via Tiepolo 11, I-34131
Trieste, Italy 
}

\PACSes{
\PACSit{98.70.Rz}{gamma-ray bursts}
\PACSit{97.60.Bw}{supernovae}
\PACSit{26.30.+k}{Nucleosynthesis in supernovae}
}

\begin{document}

\maketitle


\begin{abstract}

The connection between the long Gamma Ray Bursts (GRBs) and Type Ic
Supernovae (SNe) has revealed interesting diversity.  We review the
following types of the GRB-SN connection.  (1) GRB-SNe: The three SNe
all explode with energies much larger than those of typical SNe, thus
being called Hypernovae (HNe).  They are massive enough for forming
black holes.  (2) Non-GRB HNe/SNe: Some HNe are not associated with
GRBs.  (3) XRF-SN: SN 2006aj associated with X-Ray Flash 060218 is
dimmer than GRB-SNe and has very weak oxygen lines.  Its progenitor
mass is estimated to be small enough to form a neutron star rather
than a black hole.  (4) Non-SN GRB: Two nearby long GRBs were not
associated SNe.  Such ``dark HNe'' have been predicted in this talk
(i.e., just before the discoveries) in order to explain the origin of
C-rich (hyper) metal-poor stars.  This would be an important
confirmation of the Hypernova-First Star connection.  We will show our
attempt to explain the diversity in a unified manner with the
jet-induced explosion model.

\end{abstract}

\begin{center}
{\bf To be published in the proceedings of the conference ``SWIFT and
     GRBs: Unveiling the Relativistic Universe'',
     Venice, June 5-9, 2006. To appear in ``Il Nuovo Cimento''}
\end{center}


\section{Introduction}

\begin{figure}[t]
\begin{center}
\includegraphics*[width=9.6cm]{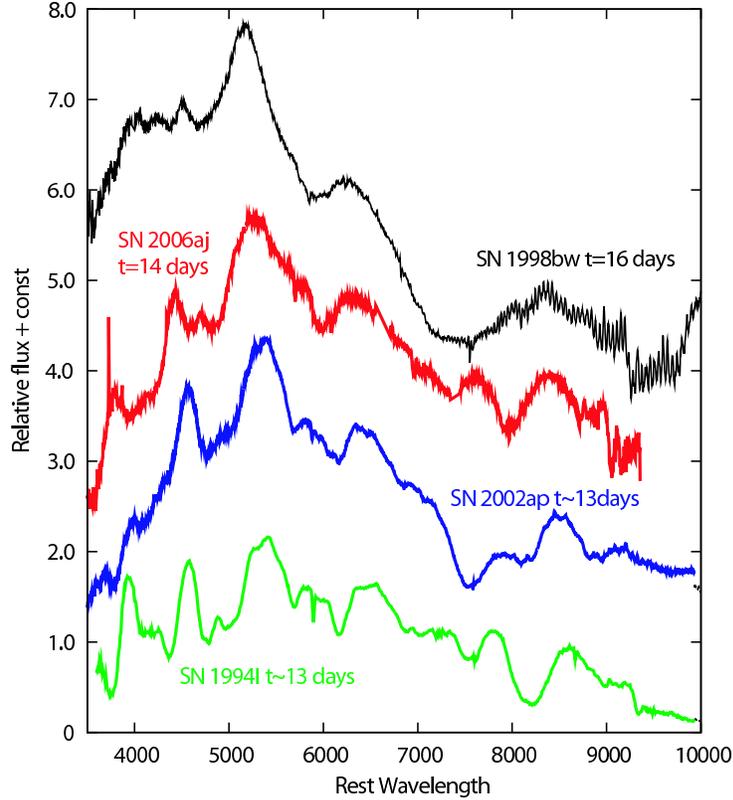}
\end{center}
\caption{
The spectra of 3 Hypernovae and 1 normal SN a few days before maximum.
SN~1998bw/GRB~980425 represents the GRB-SNe. SN~2002ap is
a non-GRB Hypernova.  SN~2006aj is associated with XRF~060218,
being similar to SN~2002ap. SN~1994I represents normal SNe.
}
\label{figSP}
\end{figure}

Long-duration $\gamma$-ray bursts (GRBs) at sufficiently close
distances ($z<0.2$) have been found to be accompanied by luminous
core-collapse Type Ic supernovae (SNe Ic) (GRB~980425/SN~1998bw
\cite{gal98}; GRB~030329/SN~2003dh \cite{sta03, hjo03};
GRB~031203/SN~2003lw \cite{mal04}).  Such GRB-SN connection is now
revealing quite a large diversity as follows.

(1) GRB-SNe: The three SNe Ic associated with the above GRBs have
similar properties; showing broader lines than normal SNe Ic
(Fig. \ref{figSP}: so-called broad-lined SNe \cite{woo06, mod07}).
These three GRB-SNe have been all found to be Hypernovae (HNe), i.e.,
very energetic supernovae, whose {\sl isotropic} kinetic energy (KE)
exceeds $10^{52}$\,erg, about 10 times the KE of normal core-collapse
SNe (hereafter $E_{51} = E/10^{51}$\,erg)
\cite{iwa98,nomoto2001,nom04}.

(2) Non-GRB HNe/SNe: These SNe show broad line features but are not
associated with GRBs (SN~1997ef \cite{iwa00}; SN~2002ap \cite{maz02};
SN 2003jd \cite{maz05}).  These are either less energetic than
GRB-SNe, or observed off-axis.

(3) XRF-SNe: X-Ray Flash (XRF) 060218 has been found to be connected
to SN Ic 2006aj \cite{campana2006, pian2006, sod06}.  The progenitor's mass
is estimated to be small enough to form a ``neutron
star-making SN'' \cite{mazzali2006b}.

\begin{figure}[t]
\begin{center}
\includegraphics[width=9.0cm]{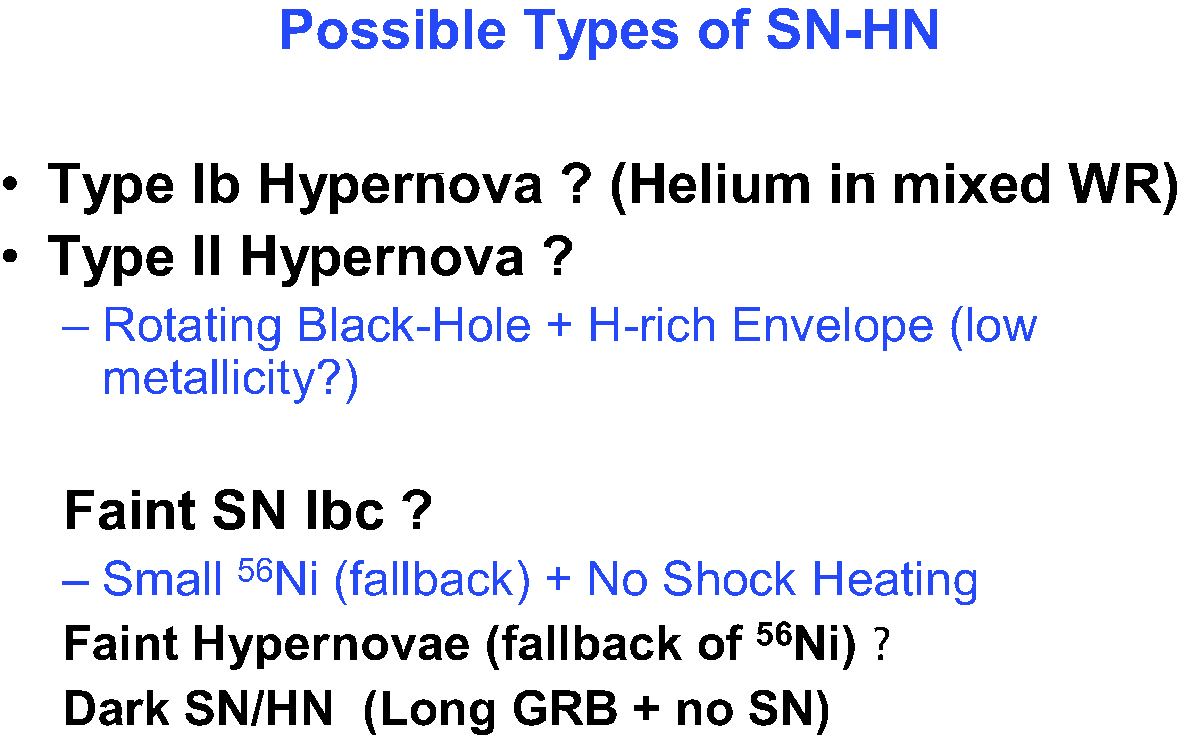}
\end{center}
\caption{The summary slide of this talk \cite{nom06b} posted on the Conference
program page at
http://www.merate.mi.astro.it/docM/OAB/Research/SWIFT/sanservolo2006/.
A Dark SN/HN (long GRB with no SN) is predicted.
}
\label{summary}
\end{figure}

(4) Non-SN GRBs: In this review talk (Fig. \ref{summary};
\cite{nom06b}), we pointed out that nucleosynthesis in HNe can explain
some of the peculiar abundance patterns (such as the large Zn/Fe and
Co/Fe ratios) in extremely metal-poor stars, which have long been
mysteries.  In particular, we have predicted that the ``dark HN'' ($=$
``non-SN GRB'' $=$ long GRB with no SN) should exist and be
responsible for the formation of the carbon-rich extremely (and hyper)
metal-poor stars (see Fig. \ref{summary}).  After the Venice
conference, the predicted ``non-SN GRBs'' have been actually
discovered (GRB 060605 and 060614) \cite{fyn06,gal06,del06,geh06}.

\begin{figure}[t]
\begin{center}
\includegraphics[width=10.0cm]{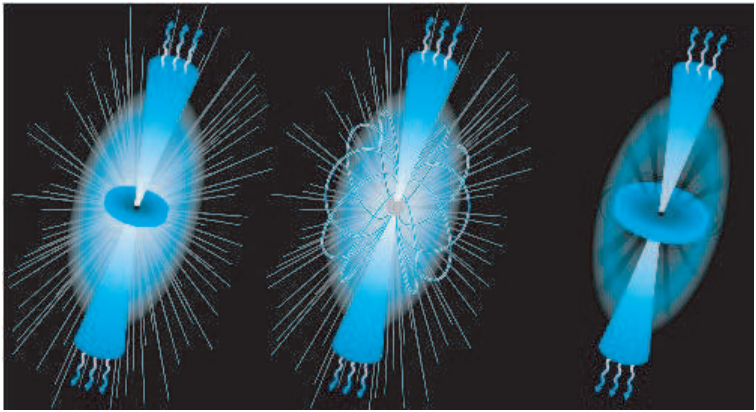}
\end{center}
\caption{Illustrations of GRB-HNe, XRF-SNe, and no-SN-GRBs (from left to
 right).}
\label{figGRBs}
\end{figure}

Here we summarize the properties and diversities of SNe connected to
the above various types of GRBs as illustrated in Figure
\ref{figGRBs}.  The model parameters are summarized in Table
\ref{tab:model}.  We then point out the GRB/Hypernova - First
Star connection through nucleosynthesis approach \cite{nomoto2006}.

\section{GRBs}

Figure \ref{figSP} compares the spectra of GRB-HNe (SN~1998bw),
non-GRB SN, XRF-SNe, and normal SN Ic.  The spectrum of SN~1998bw has
very broad lines.  The strongest absorptions are \TiII-\FeII\
(shortwards of $\sim 4000$\AA, \FeII-\FeIII\ (near 4500\AA), \SiII\
(near 5700\AA), and \OI-\CaII\ (between 7000 and 8000 \AA).  We
calculate the synthetic spectra for ejecta models of bare C+O stars
with various ejected mass $\Mej$ and $E$.  The large $E$/$\Mej$ is
required to reproduce the broad features.

The spectroscopic modelings are combined with the light curve (LC)
modeling as seen in Figure \ref{figLCfit} to give the estimates of
$\Mej$ and $E$.  The timescale of the LC around maximum brightness
reflects the timescale for optical photons to diffuse \cite{arn82}.
For larger $\Mej$ and smaller $E$, the LC peaks later and the LC width
becomes broader because it is more difficult for photons to escape.

 From the synthetic spectra and light curves, it was interpreted as
the explosion of a massive star, with $E_{51} \sim 30$
and $\Mej \sim 10 \Msun$\cite{iwa98}.  Also the very high luminosity of
SN 1998bw indicates that a large amount of \Nifs\ ($\sim 0.5 \Msun$)
was synthesized in the explosion.

The ejected \Nifs\ mass is estimated to be $\Mni\sim0.3-0.7\Msun$ (\eg
\cite{maz06a}) which is 4 to 10 times larger than typical SNe Ic
($\Mni\sim 0.07\Msun$ \cite{nomoto2006}).

The other two GRB-SNe, 2003dh and 2003lw, are also characterized by
the very broad line features and the very high luminosity.  $\Mej$ and
$E$ are estimated from synthetic spectra and light curves and
summarized in Figure \ref{figME}\cite{nak01a, deng05, maz06a}.  It is
clearly seen that GRB-SNe are the explosions of massive progenitor
stars (with the main sequence mass of $M_{\rm ms} \sim 35 - 50
\Msun$), have large explosion kinetic energies ($E_{51} \sim 30 -
50$), synthesized large amounts of \Nifs\ ($\sim 0.3 - 0.5 \Msun$).

These GRB-associated HNe (GRB-HNe) are suggested to be the outcome
of very energetic black hole (BH) forming explosions of massive stars
(\eg \cite{iwa98}). 

Hypernovae are also characterized by asphericity from the observations
of polarization and emission line features \cite{wang2003, kaw02,
mae02}.  The explosion energy of the aspherical models for hypernovae
tends to be smaller than the spherical models by a factor of 2 - 3,
but still being as high as $E_{51} \gsim 10$
\cite{mae06LC}.

\begin{table}
  \caption{The models for Type Ic SNe and HNe (masses $M$ are in unit of
  $M_{\odot}$ and $E_{51}=E/10^{51}$ ergs).}
  \label{tab:model}
  \begin{tabular}{ccccccccc}
   \hline
    Name & $M_{\rm ms}$ & $M_{\rm ej}$ & $E_{51}$ & $M({\rm ^{56}Ni})$ &
    $E_{51}/M_{\rm ej}$ & $M(v>30,000{\rm km~s^{-1}})$ \\
   \hline
   SN~1998bw & 40   & 10.4 & 40    & 0.4  & 3.8  & 1.46 \\
   SN~2003dh & 32.5 &  7   & 35    & 0.4  & 5    & 1.41 \\
   SN~2003lw & 55   & 13   & 60    & 0.55 & 4.2  & 1.76 \\
\\
   SN~1997ef & 32.5 & 8.6  & 12.75 & 0.15 & 1.5  & 0.36 \\
\\
   SN~2002ap & 22.5 & 3.3  & 4     & 0.07 & 1.2  & 0.055 \\
   SN~2006aj & 20   & 1.8  & 2     & 0.21 & 1.1  & 0.019 \\
\\
   SN~1994I  & 14   & 1    & 1     & 0.07 & 1    & 0.0028 \\ 
   \hline
  \end{tabular}
\end{table}

\section{Non-GRB Hypernovae/Supernovae}

These HNe show spectral features similar to those of GRB-SNe but are
not known to have been accompanied by a GRB.  The estimated $\Mej$ and
$E$, obtained from synthetic light curves and spectra, show that there
is a tendency for non-GRB HNe to have smaller $\Mej$, $E$, and lower
luminosities as summarized in Table \ref{tab:model} and Figure
\ref{figME}.

SN 1997ef is found to be the HN class of energetic explosion, although
$E/M_{\rm ej}$ is a factor 3 smaller than GRB-SNe
(Fig. \ref{figM30000}).  It is not clear whether SN 1997ef is not
associated with GRB because of this smaller $E/M_{\rm ej}$ or it was
actually associated with the candidate GRB 971115.

SN 2002ap was not associated a GRB and no radio has been observed.  It
has similar spectral features, but narrower and redder
(Fig. \ref{figSP}), which was modeled as a smaller energy explosion,
with $E_{51} \sim 4$ and $\Mej \sim 3
\Msun$\cite{maz02}.

The early time spectrum of SN 2003jd is similar to SN 2002ap.
Interestingly, the its nebular spectra shows a double peak in
O-emission lines \cite{maz05}.  This has exactly confirmed the
theoretical prediction by the asymmetric explosion model \cite{mae02}.
In this case, the orientation effect might cause the non-detection of
a GRB (see Mazzali et al. in this volume for more details.)

Whether the non-detection of GRBs is the effect of different
orientations or of an intrinsic property is still a matter of debate,
but there is a tendency for them to have smaller $\Mej$ and $E$.  SN
1997ef seems to be a marginal case.

\section{XRF--Supernovae}

GRB060218 is the second closest event as ever ($\sim 140\,$Mpc).  The
GRB was weak \cite{campana2006} and classified as X-Ray Flash (XRF)
because of its soft spectrum.  The presence of SN 2006aj was soon
confirmed\cite{pian2006,mod06}.  Here we summarize the properties of
SN 2006aj by comparing with other SNe~Ic.

\begin{figure}[t]
\begin{center}
\includegraphics[width=9.4cm]{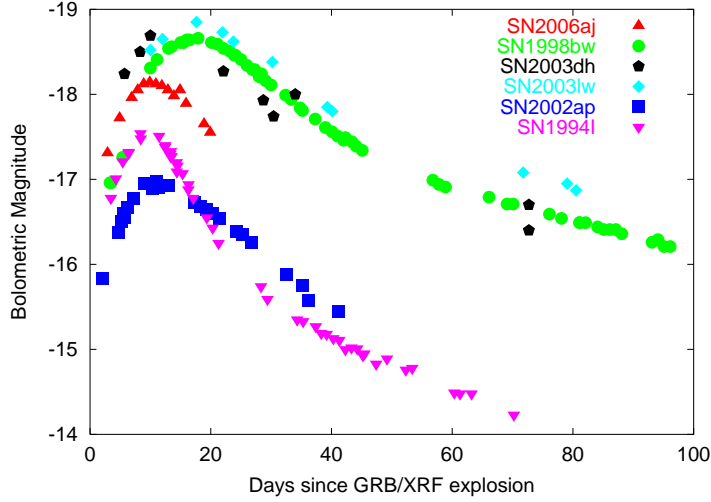}
\end{center}
\caption{
The bolometric light curves of GRB-SN (SNe 1998bw, 2003dh), non-GRB-SN
(2002ap), XRF-SN (2006aj), and normal SNe Ic (1994I) are compared.}
\label{figLC}
\end{figure}

\begin{figure}[t]
\begin{center}
\includegraphics[width=9.4cm]{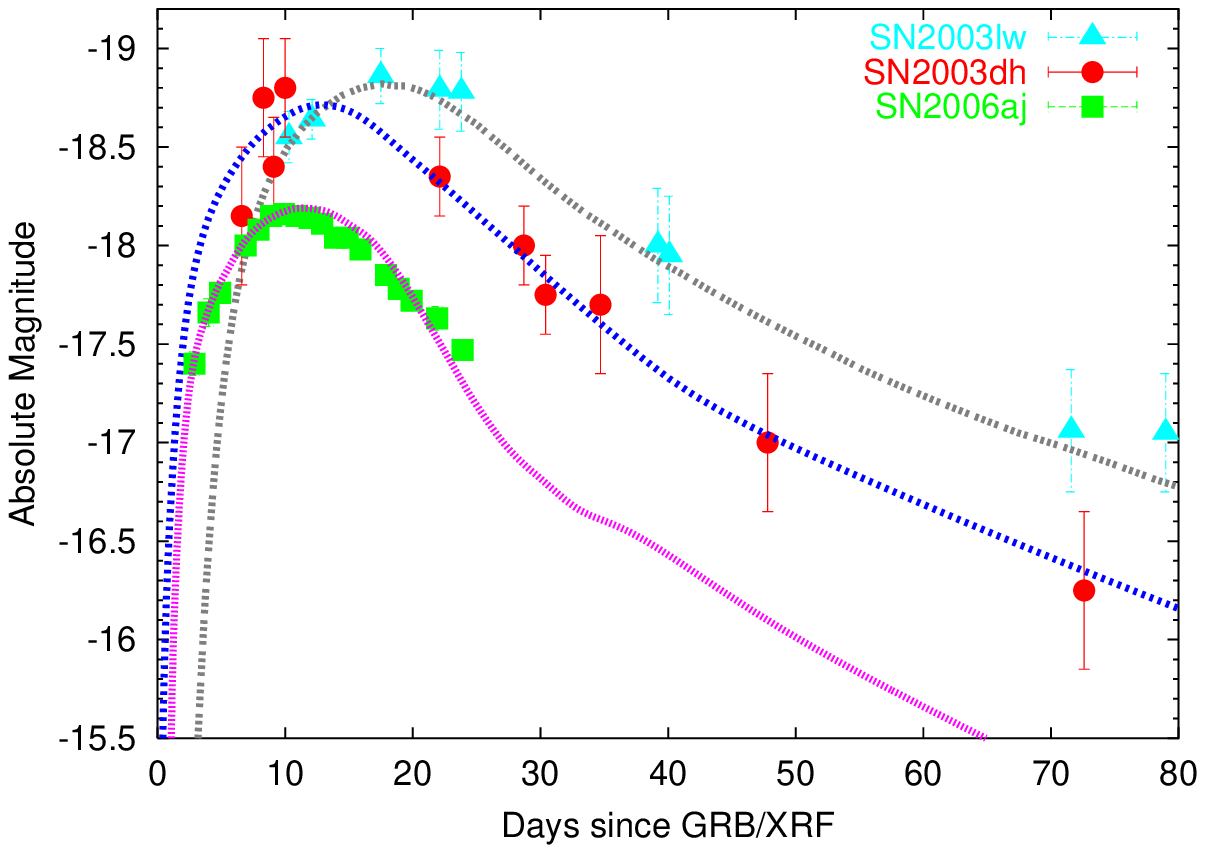}
\end{center}
\caption{
Theoretical bolometric light curves for SNe 2003dh, 2003lw, and 2006aj.
}
\label{figLCfit}
\end{figure}

SN~2006aj has several features that make it unique.  It is less bright
than the other GRB/SNe (Fig. \ref{figLC}).  Its rapid photometric
evolution is very similar to that of a dimmer, non-GRB SN
2002ap\cite{maz02}, but it is somewhat faster.  Although its spectrum
is characterized by broader absorption lines as in SN 1998bw and other
GRB/SN, they are not as broad as those of SN~1998bw, and it is much
more similar to that of SN~2002ap (Fig. \ref{figSP}).  The most
interesting property of SN~2006aj is surprisingly weak oxygen lines,
much weaker than in Type Ic SNe.

\subsection{Spectroscopic and Photometric Models}

\begin{figure}[t]
\begin{center}
\includegraphics[width=9.0cm]{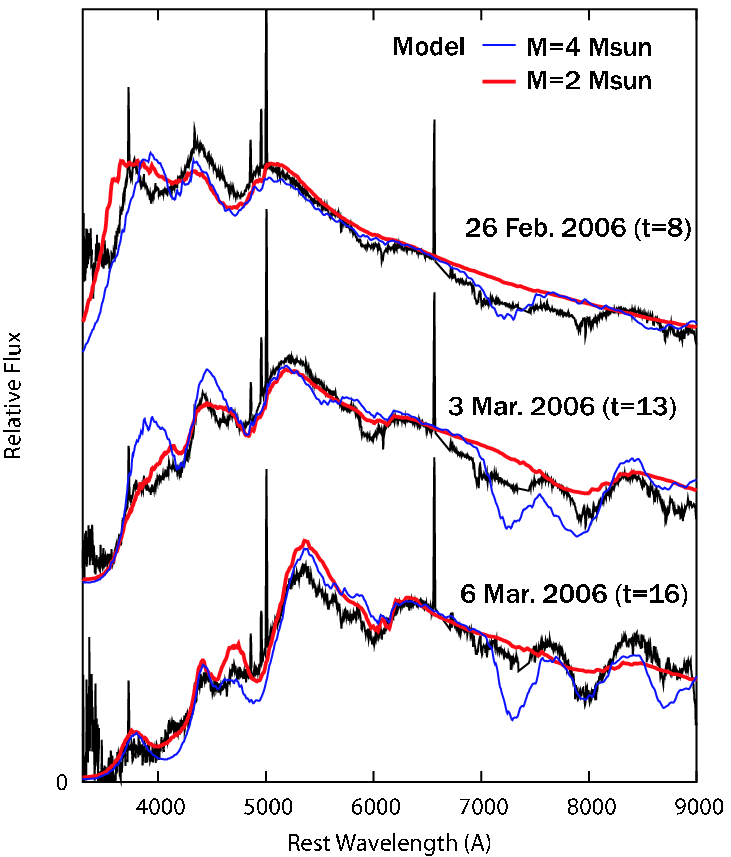}
\end{center}
\caption{Synthetic spectra with a explosion model with 
$(\Mej, E_{51})$ = $(2.0 \Msun , 2.0)$ [gray solid lines]
and $(4.0 \Msun , 9.0)$ [black dashed lines]
compared with the observed spectra of SN 2006aj (solid lines).}
\label{figspfit}
\end{figure}

By modeling the spectra and the light curve, we derive for SN~2006aj
$\Mej \sim 2 \Msun$ and $E_{51} \sim 2$ as follows
(Figs. \ref{figLCfit} and \ref{figspfit}).  Lack of oxygen in the
spectra indicates $\sim 1.3 \Msun$ of O, and oxygen is still the
dominant element.

The strength of the OI$\lambda$7774 line, which is the strongest
oxygen line in optical wavelength, is sensitive to the temperature in
the ejecta.  Since the fraction of OI is larger in the lower
temperature ejecta (although OII is still the dominant ionization
state), the normal SNe Ib/c always show the strong OI absorption (see
SN 1994I in Fig. \ref{figSP}) irrespective of the ejecta mass.

In more luminous SNe like GRB-SNe and SN 2006aj, the OI fraction tends
to be smaller.  However, if the ejecta are very massive, \eg $\sim 10
\Msun$, the mass of OI is large enough to make the strong absorption
(see SN 1998bw in Fig. \ref{figSP}).  In the case of SN 2006aj, the
temperature is larger than in normal SNe Ib/c.  Therefore, the weak OI
line indicates that the ejecta mass is not as massive as SN 1998bw,
which supports our conclusion.

\begin{figure}[t]
\begin{center}
\includegraphics[width=8.8cm]{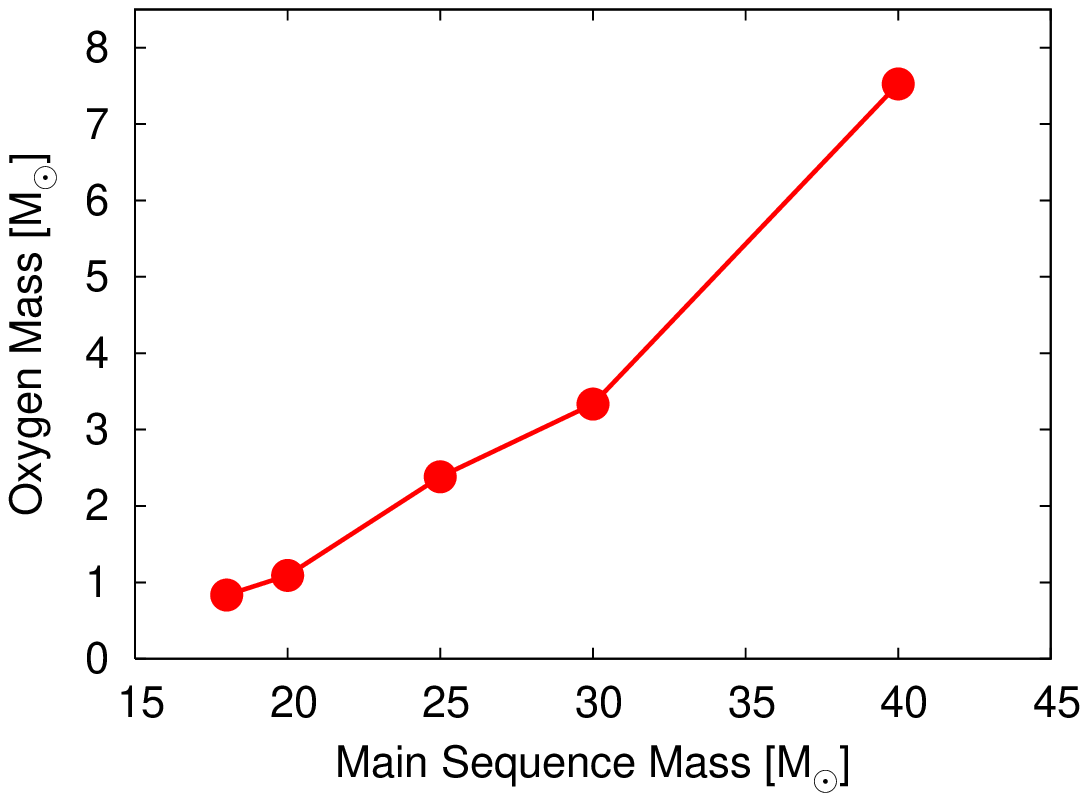}
\end{center}
\caption {The synthesized oxygen mass as a function of the 
progenitor's main-sequence mass.}
\label{figMO}
\end{figure}

The spectroscopic results are confirmed through the light curve
modeling (Fig. \ref{figLCfit}).  We synthesize the theoretical light
curve for the density and chemical abundance structure to compare with
SN~2006aj.  The best match is achieved with a total
\Nifs\ mass of $0.21 \Msun$ in which $0.02 \Msun$ is located above 
20,000\kms (Fig. \ref{figLC}).  The high-velocity \Nifs\ is
responsible for the fast rise of the light curve, because photons
created can escape more easily.  In a realistic asymmetric explosion,
such high-velocity $^{56}$Ni could abundantly be produced along the
direction of the GRB jets\cite{mae02,mae03}.

\subsection{The Progenitor and Implications for XRF}

The properties of SN~2006aj (smaller $E$ and smaller $\Mej$) suggest
that SN~2006aj is not the same type of event as the other GRB-SNe
known thus far.  One possibility is that the initial mass of the
progenitor star is much smaller than the other GRB-SNe, so that the
collapse/explosion generated less energy.  If $M_{\rm ms}$ is $\sim 20
- 25 \Msun$, the star would be at the boundary between collapse to a
black hole or to a neutron star.  In this mass range, there are
indications of a spread in both $E$ and the mass of \Nifs\
synthesized\cite{hamuy2003}.  The fact that a relatively large amount
of \Nifs\ is required in SN 2006aj possibly suggests that the star
collapsed only to a neutron star because more core material would be
available to synthesize \Nifs\ in the case.

Although the kinetic energy of $E_{51}\sim 2$ is larger
than the canonical value ($1 \times 10^{51}$ erg, \cite{nom94}) in the
mass range of $M_{\rm ms} \sim 20 - 25 \Msun$, such an energy might be
obtained from magnetar-type activity.  It is conceivable that in this
weaker explosion than typical GRB-SNe, the fraction of energy
channeled to relativistic ejecta is smaller, giving rise to an XRF
rather than a classical GRB.

Another case of a SN associated with an XRF has been reported
(XRF030723)\cite{fyn04}.  The putative SN, although its spectrum was
not observed, was best consistent with the properties of
SN~2002ap\cite{tom05}.  This may suggest that XRFs are associated with
less massive progenitor stars than those of canonical GRBs, and that
the two groups may be differentiated by the formation of a neutron
star\cite{nak98} or a BH.  Still, the progenitor star must have been
thoroughly stripped of its H and He envelopes, which is a general
property of all GRB-SNe and probably a requirement for the emission of
a high energy transient.  These facts may indicate that the progenitor
is in a binary system.

If magnetars are related to the explosion mechanism, some short
$\gamma$-ray repeaters energized by a magnetar\cite{tho95,tho04} may
be remnants of GRB060218-like events.  Magnetars could generate a GRB
at two distinct times.  As they are born, when they have a very large
spin rate ($\sim 1$ ms), an XRF (or a soft GRB) is produced as in
SN\,2006aj/GRB060218. Later (more than 1,000 yrs), when their spin
rate is much slower, they could produce short-hard GRBs \cite{hur05}.

Stars of $M_{\rm ms}\sim20-25 \Msun$ are much more common than stars
of $M_{\rm ms}\sim35-50 \Msun$, and so it is highly likely that events
such as GRB060218 are much more common in nature than the highly
energetic GRBs.  They are, however, much more difficult to detect
because they have a low $\gamma$-ray flux.  The discovery of
GRB060218/SN~2006aj suggests that there may be a wide range of
properties of both the SN and the GRB in particular in this mass
range.  The continuing study of these intriguing events will be
exciting and rewarding.

\begin{figure}[t]
\begin{center}
\includegraphics[width=9.5cm]{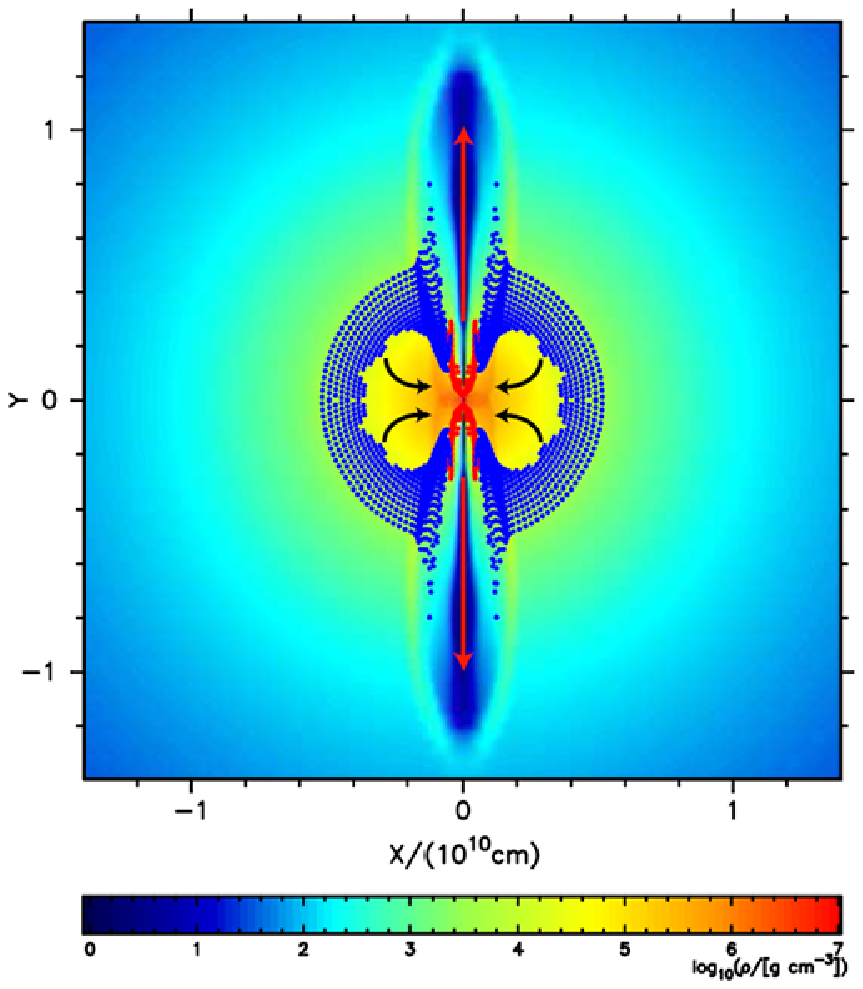}
\end{center}
\caption[f1.eps]{The density structure of the 40 $\Msun$ Pop III star
 explosion model of $\Edep=15$ at 1 sec after the start of the jet
 injection. The jets penetrate the stellar mantle ({\it red arrows}) and
 material falls on the plane perpendicular to the jets ({\it black arrows}). 
 The dots represent ejected Lagrangian elements dominated by Fe 
 (\Nifs, {\it red}) and by O ({\it blue}). 
\label{fig:fallback}}
\end{figure}

\section{Non-SN Gamma-Ray Bursts}

For recently discovered nearby long-duration GRB~060505 ($z=0.089$,
\cite{fyn06}) and GRB~060614 ($z=0.125$, \cite{gal06,fyn06,del06,geh06}), 
no SN has been detected.  Upper limits to brightness of the possible
SNe are about 100 times fainter than SN~1998bw. These correspond to
upper limits to the ejected \Nifs\ mass of $\Mni\sim 10^{-3}\Msun$.

A small amount of \Nifs\ ejection has been indicated in the faintness
of several Type II SNe (SNe II, \eg SN~1994W, \cite{sollerman1998};
and SN~1997D, \cite{turatto1998}). The estimated $E$ of these faint
SNe~II are very small ($E_{\rm 51}\lsim1$,
\cite{turatto1998}). These properties are well-reproduced by the {\sl
spherical} explosion models that undergo a large amount of fallback if
$E$ is sufficiently small \cite{woo95,iwa05}. Thus these faint SN
explosions with low $E$ seem to be superficially irreconcilable to the
formation of energetic GRBs \cite{gal06}.  We will show that the
existence of a high-energy narrow jet which produces a GRB is
compatible with a faint/dark and low $E$ SN with little \Nifs\
ejection.

Tominaga et al. \cite{tom07} calculated the jet-induced explosions
(\eg \cite{mae03,nag06}) of the $40\Msun$ stars
\cite{ume05,tominaga2006} by injecting the jets at a radius 
$R \sim 900$ km, corresponding to an enclosed mass of $M \sim 1.4
\Msun$.  They investigated the dependence of nucleosynthesis outcome
on $\Ed$ for a range of $\Edep\equiv\Ed/10^{51}{\rm
ergs\,s^{-1}}=0.3-1500$.  The diversity of $\Ed$ is consistent with
the wide range of the observed isotropic equivalent $\gamma$-ray
energies and timescales of GRBs (\cite{ama06} and references therein).
Variations of activities of the central engines, possibly
corresponding to different rotational velocities or magnetic fields,
may well produce the variation of $\Ed$.

After the jet injection is initiated, the shock fronts between the
jets and the infalling material proceed outward in the stellar
mantle.  Figure~\ref{fig:fallback} is a snapshot of the model with
$\Edep=15$ at 1 sec after the start of jet injection.  When the jet
injection ends, the jets have been decelerated by collisions with the
dense stellar mantle and the shock has become more spherical.  The
inner materials are ejected along the jet-axis but not along the
equatorial plane.  

\begin{figure}[t]
\begin{center}
\includegraphics[width=9.5cm]{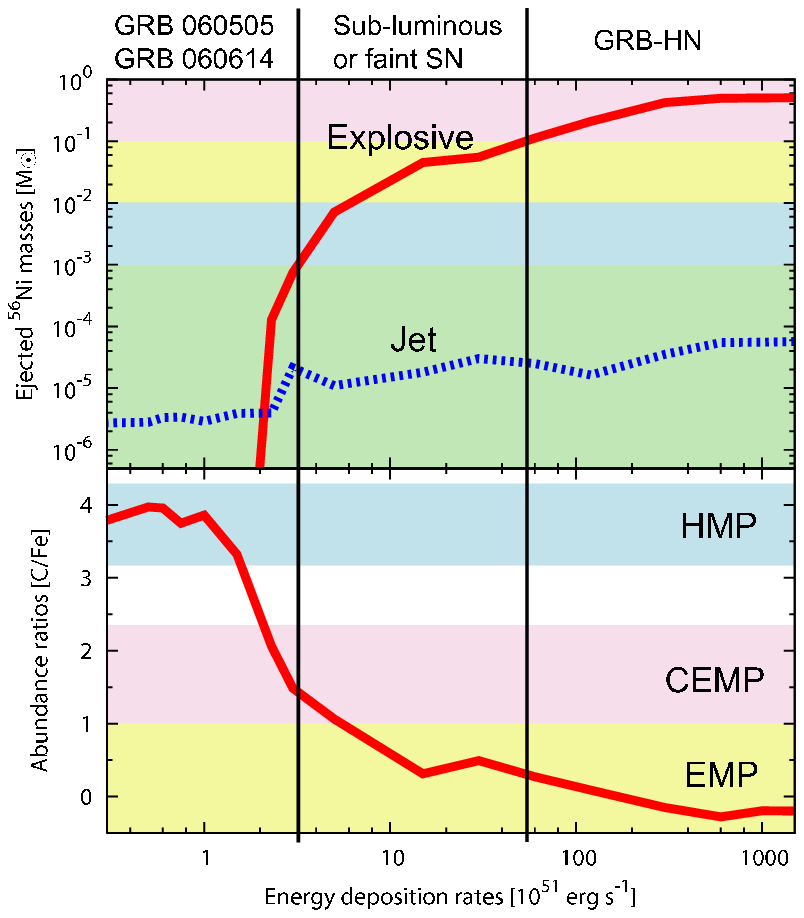}
\end{center}
\caption{{\it Top}: the ejected \Nifs\ mass ({\it red}: 
 explosive nucleosynthesis products, {\it blue}: the jet contribution)
 as a function of the energy deposition rate. The background color
 shows the corresponding SNe ({\it red}: GRB-HNe, {\it yellow}:
 sub-luminous SNe, {\it blue}: faint SNe, {\it green}: GRBs~060505 and
 060614).  Vertical lines divide the resulting SNe according to their
 brightness.  {\it Bottom}: the dependence of abundance ratio [C/Fe]
 on the energy deposition rate. The background color shows the
 corresponding metal-poor stars ({\it yellow}: EMP, {\it red}: CEMP,
 {\it blue}: HMP stars).
\label{fig:EdotNi}}
\end{figure}

\subsection{\Nifs\ Mass}

The top panel of Figure~\ref{fig:EdotNi} shows the dependence of the
ejected $\Mni$ on the energy deposition rate $\Ed$. For lower $\Ed$,
smaller $\Mni$ is synthesized in explosive nucleosynthesis because of
lower post-shock densities and temperatures (\eg
\cite{mae03,nag06,mae06c}).  

If $\Edep\gsim3$, the jet injection is initiated near the bottom of
the C+O layer, leading to the synthesis of $\Mni\gsim10^{-3}\Msun$. If
$\Edep<3$, on the other hand, the jet injection is delayed and
initiated near the surface of the C+O core; then the ejected \Nifs\ is
as small as $\Mni<10^{-3}\Msun$.

\Nifs\ contained in the relativistic jets is only
$\Mni\sim10^{-6}-10^{-4}\Msun$ because the total mass of the jets is
$M_{\rm jet}\sim 10^{-4}\Msun$ in our model with ${\it \Gamma}_{\rm
jet} = 100$ and $E_{\rm dep}=1.5\times10^{52}$ergs.  Thus the \Nifs\
production in the jets dominates over explosive nucleosynthesis in the
stellar mantle only for $\Edep<1.5$ in the present model.

\begin{figure}[t]
\begin{center}
\includegraphics[width=10.4cm]{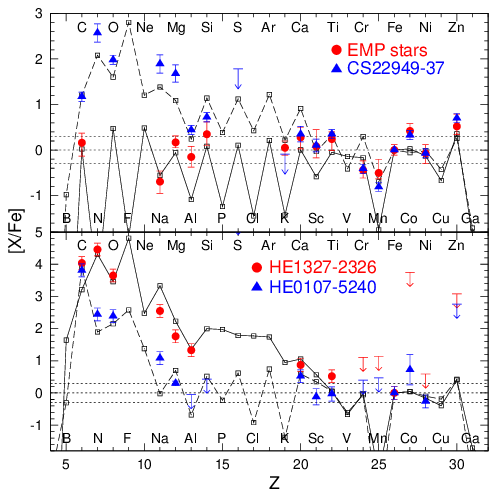}
\end{center}
\caption{A comparison of the abundance patterns of metal-poor 
 stars and of our models. 
 {\it Top}: typical EMP ({\it red dots}, \cite{cayrel2004}) and 
 CEMP ({\it blue triangles}, CS~22949--37, \cite{dep02}) stars and models
 with $\Edep=120$ ({\it solid line}) and $=3.0$ ({\it dashed line}).
 {\it Bottom}: HMP stars:
 HE~1327--2326, ({\it red dots}, \eg \cite{fre05}), 
 and HE~0107--5240, ({\it blue triangles}, \cite{chr02,bes05}) and models 
 with $\Edep=1.5$ ({\it solid line}) and $=0.5$ ({\it dashed line}).
\label{fig:EMP}}
\end{figure}

\subsection{GRB-HNe}

For high energy deposition rates ($\Edep\gsim60$), the explosions
synthesize large $\Mni$ ($\gsim0.1\Msun$) being consistent with
GRB-HNe.  The remnant mass was $M_{\rm rem}^{\rm start}\sim1.5\Msun$
when the jet injection was started, but it grows as material is
accreted from the equatorial plane. The final BH masses range from
$M_{\rm BH}=10.8\Msun$ for $\Edep=60$ to $M_{\rm BH}=5.5\Msun$ for
$\Edep=1500$, which are consistent with the observed masses of
stellar-mass BHs \cite{bai98}.  The model with $\Edep=300$ synthesizes
$\Mni\sim0.4\Msun$ and the final mass of BH left after the explosion
is $M_{\rm BH}=6.4\Msun$.

\subsection{Non-SN GRBs (060505 and 060614)}

For low energy deposition rates ($\Edep<3$), the ejected
\Nifs\ masses ($\Mni<10^{-3}\Msun$) are smaller than the upper
limits for GRBs~060505 and 060614. The final BH mass is larger for
smaller $\Ed$.
While the material ejected along the jet-direction involves those from
the C+O core, the material along the equatorial plane fall back.

If the explosion is viewed from the jet direction, we would observe
GRB without SN re-brightening. This may be the situation for
GRBs~060505 and 060614.  In particular, for $\Edep<1.5$, \Nifs\ cannot
be synthesized explosively and the jet component of the Fe-peak
elements dominates the total yields (Fig.~\ref{fig:EdotNi}). The
models eject very little $\Mni$ ($\sim10^{-6}\Msun$).

\subsection{GRBs with Faint or Sub-Luminous SNe}

For intermediate energy deposition rates ($3\lsim\Edep<60$), the
explosions eject $10^{-3}\Msun \lsim \Mni <0.1\Msun$ and the final BH
masses are $10.8\Msun\lsim M_{\rm BH}< 15.1\Msun$. The resulting SN is
faint ($\Mni <0.01\Msun$) or sub-luminous ($0.01\Msun \lsim \Mni
<0.1\Msun$).

Nearby GRBs with faint or sub-luminous SNe have not been observed.
This may be because they do not occur intrinsically in our
neighborhood or because the number of observed cases is still too
small. In the latter case, further observations may detect GRBs with a
faint or sub-luminous SN.

\subsection{Abundance Patters of C-rich Metal-Poor Stars}

The bottom panel of Figure~\ref{fig:EdotNi} shows the dependence of
the abundance ratio [C/Fe] on $\Ed$. Lower $\Ed$ yields larger $M_{\rm
BH}$ and thus larger [C/Fe], because the infall decreases the amount
of inner core material (Fe) relative to that of outer material (C)
(see also \cite{mae03}). As in the case of $\Mni$, [C/Fe] changes
dramatically at $\Edep\sim3$.

The abundance patterns of the EMP stars are good indicators of SN
nucleosynthesis because the Galaxy was effectively unmixed at [Fe/H]
$< -3$ (\eg \cite{tum06}). They are classified into three groups
according to [C/Fe]:

(1) [C/Fe] $\sim 0$, normal EMP stars ($-4<$ [Fe/H] $<-3$, \eg
    \cite{cayrel2004});

(2) [C/Fe] $\gsim+1$, Carbon-enhanced EMP (CEMP) stars ($-4<$ [Fe/H]
    $<-3$, \eg CS~22949--37 \cite{dep02}); 

(3) [C/Fe] $\sim +4$, hyper metal-poor (HMP) stars ([Fe/H] $<-5$, \eg
    HE~0107--5240 \cite{chr02,bes05}; HE~1327--2326 \cite{fre05}).

Figure \ref{fig:EMP} shows that the general abundance patterns of the
normal EMP stars, the CEMP star CS~22949--37, and the HMP stars
HE~0107--5240 and HE~1327--2326 are reproduced by models with
$\Edep=120$, 3.0, 1.5, and 0.5, respectively (see Table~1 for model
parameters). The model for the normal EMP stars ejects
$\Mni\sim0.2\Msun$, i.e. a factor of 2 less than SN~1998bw. On the
other hand, the models for the CEMP and the HMP stars eject
$\Mni\sim8\times10^{-4}\Msun$ and $4\times 10^{-6}\Msun$,
respectively, which are always smaller than the upper limits for
GRBs~060505 and 060614.

\begin{figure}[t]
\begin{center}
\includegraphics[width=10.2cm]{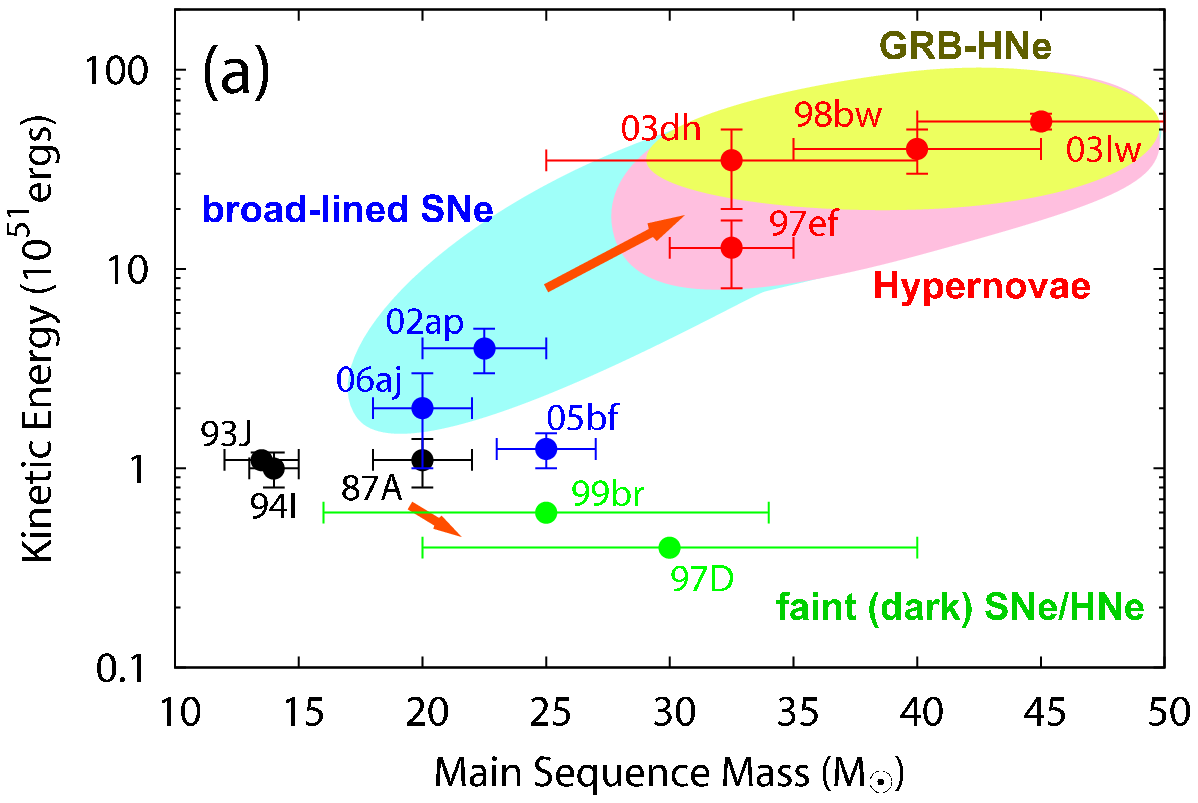}\\
\includegraphics[width=10.2cm]{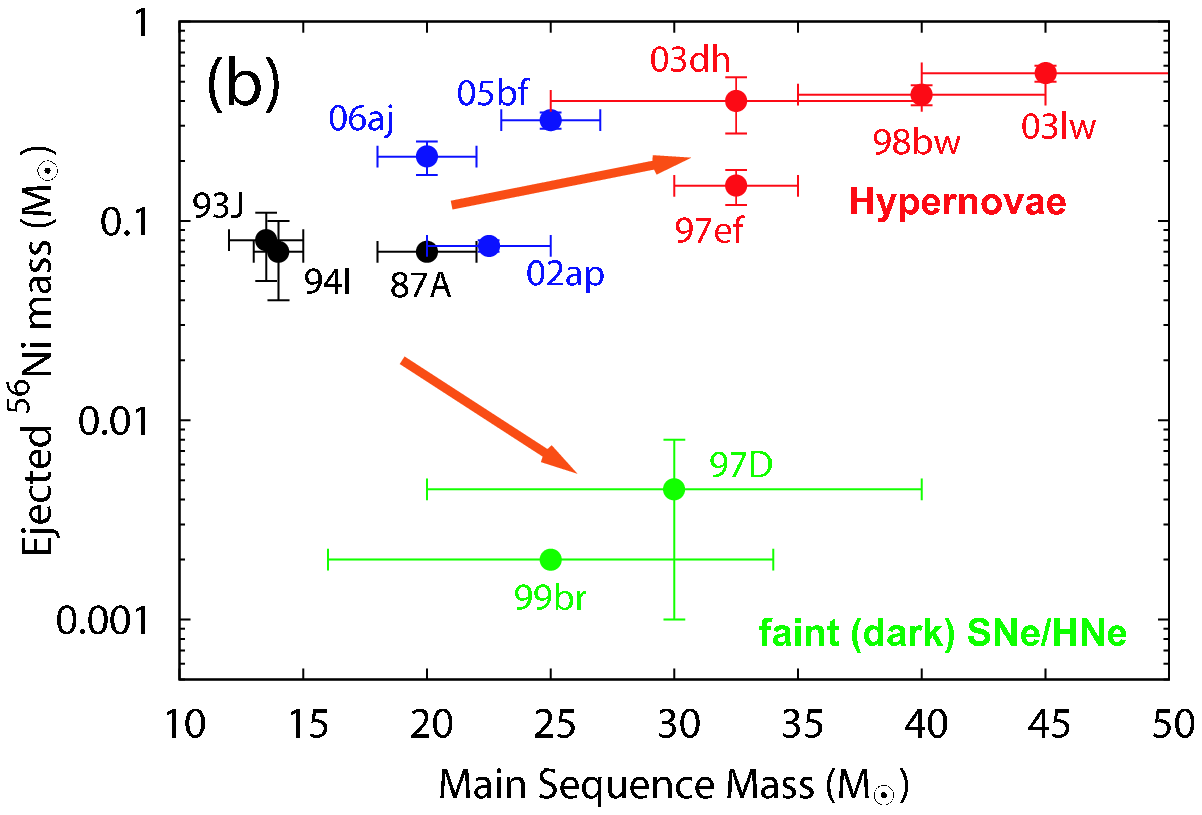}
\end{center}
\caption{
The kinetic explosion energy $E$ and the ejected $^{56}$Ni mass as a
function of the main sequence mass $M$ of the progenitors for several
supernovae/hypernovae. SNe that are observed to show broad-line features
are indicated.  Hypernovae are the SNe with $E_{51}>10$. 
}
\label{figME}
\end{figure}

\section{Discussion}

Figure \ref{figME} summarize the properties of core-collapse SNe as a
function of the main-sequence mass $M_{\rm ms}$ of the progenitor
star \cite{nomoto2003}.  Figure \ref{figM30000} shows how the properties of these SNe
depend on $E/\Mej$ and the mass contained at the the expansion
velocity exceeding $v =$ 30,000 km s$^{-1}$.

\subsection{GRB, Hypernovae, Broad-Line features}

The broad-line SNe include both GRB-SNe and Non-GRB SNe.

\begin{figure}[t]
\begin{center}
\includegraphics[width=9.3cm]{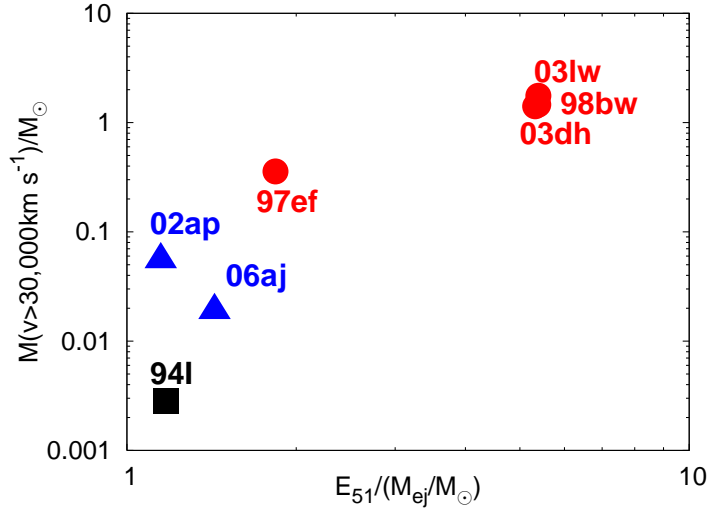}
\end{center}
\caption{
The mass (in unit of $M_\odot$) contained in the outer layer of the
models where the expansion velocities exceed $v=$ 30,000 km s$^{-1}$
against $E_{51}/(\Mej/M_\odot)$.
}
\label{figM30000}
\end{figure}

(1) GRB vs. Non-GRB: Three GRB-SNe are all similar Hypernovae (i.e.,
$E_{51} \gsim$ 10.  Thus $E$ could be closely related to
the formation of GRBs.  SN 1997ef seems to be a marginal case.  As
seen in Figure \ref{figM30000}, $E/\Mej$ could be more important
because SN 1997ef has significantly smaller $E/\Mej$ than GRB-SNe.

(2) Broad-Line features: The mass contained at $v >$ 30,000 km
s$^{-1}$ (or even higher boundary velocity) might be critical in
forming the broad-line features, although further modeling is required
to clarify this point.

\subsection{XRFs, GRBs, and SNe Ibc from the 20 - 25 $\Msun$ Progenitors}

The discovery of XRF~060218/SN~2006aj and their properties extend the
GRB-HN connection to XRFs and to the HN progenitor mass as low as
$\sim 20 \Msun$.  The XRF~060218 may be driven by a neutron star
rather than a black hole.

The final fate of 20 - 25 $\Msun$ stars show interesting variety.
Even normal SN Ib 2005bf is very different from previously known
SNe/HNe \cite{tom05b,fol05}.  This mass range corresponds to the
transition from the NS formation to the BH formation.  The NSs from
this mass range could be much more active than those from lower mass
range because of possibly much larger NS masses (near the maximum
mass) or possibly large magnetic field (i.e., Magnetar).  XRFs and
GRBs from this mass range of 20 - 25 $\Msun$ might form a different
population.

\subsection{Hypernovae of Type II and Type Ib?}

Suppose that smaller losses of mass and angular momentum from low
metallicity massive stars lead to the formation of more rapidly
rotating NSs or BHs and thus more energetic explosions.  Then we
predict the existence of Type Ib and Type II HNe \cite{ham05}, as
summarized in the slide of Figure \ref{summary} \cite{nom06b}

So far all observed HNe are of Type Ic.  However, most of SNe Ic are
suggested to have some He \cite{bra06}.  If even the small amount of
radioactive $^{56}$Ni is mixed in the He layer, the He feature should
be seen \cite{luc91,nom95}.  For HNe, the upper mass limit of He has
been estimated to be $\sim 2 \Msun$ \cite{maz02} for the case of no He
mixing.  If He features would be seen in future HN observations, it
would provide an important constraint on the models, especially, the
fully mixed WR models \cite{yoo05,woo06b,mey07}.

\subsection{Hypernova-First Star Connection}

Based on the results in the earlier section, we suggest that the first
generation supernovae were the explosion of $\sim$ 20-130 $M_\odot$
stars and some of them produced C-rich, Fe-poor ejecta.

We have computed hydrodynamics and nucleosynthesis for the explosions
induced by relativistic jets. We have shown that (1) the explosions
with large $\Ed$ are observed as GRB-HNe and their yields can explain
the abundances of normal EMP stars, and (2) the explosions with small
$\Ed$ are observed as GRBs without bright SNe and can be responsible
for the formation of the CEMP and the HMP stars. We thus propose that
GRB-HNe and GRBs without bright SNe belong to a continuous series of
BH-forming massive stellar deaths with the relativistic jets of
different $\Ed$.

A short GRB, probably the result of the merger of two compact objects
(\eg \cite{geh05}), synthesizes virtually no \Nifs\ because the ejecta
must be too neutron-rich.  In contrast, our model suggests that
GRBs~060505 and 060614 produced $\Mni\sim10^{-4}-10^{-3}\Msun$ or
$\sim 10^{-6}\Msun$.  If such a GRB without a bright SN occurs in a
very faint and nearby galaxy, our model predicts that some
re-brightening due to the \Nifs\ decay can be observed.

The nearby GRB-HNe and GRBs without bright SNe eject
$\Mni\sim0.3-0.7\Msun$ and $<10^{-3}\Msun$, respectively.  The
GRB-associated faint or sub-luminous SNe with
$10^{-3}\Msun\lsim\Mni<0.3\Msun$ have never been observed at close
distances.  Possible reasons may be that (1) they do not occur
intrinsically, \ie the energy deposition rate is bimodally
distributed, or that (2) the number of observed nearby GRBs is still
too small. For case (1), the GRB progenitors may be divided into two
groups, \eg with rapid or slow rotation and/or with strong or weak
magnetic fields. For case (2), future observations will detect GRBs
associated with a faint or sub-luminous SN.


\acknowledgments

This work has been supported in part by the Grant-in-Aid for
Scientific Research (17030005, 17033002, 18104003, 18540231 for K.N.)
and the 21st Century COE Program (QUEST) from the JSPS and MEXT of
Japan.


\end{document}